\documentclass[mathleft
]{an}
\usepackage{graphicx}
\usepackage{times}
\newcommand{\Ms}{$M_{\star}$}

\begin{document}

% The following seven commands are intended for editorial usage and should be ignored by
% the author(s).
\Pagespan{789}{}% Document's page range. 
% If second parameter is left empty, the last page is computed automatically.
\Yearpublication{2012}%
\Yearsubmission{2012}%
\Month{10}%   
\Volume{ }%  
\Issue{ }% 
% \DOI{This.is/not.aDOI}% 

\title{Galaxy And Mass Assembly (GAMA): The M-Z relation for galaxy groups}

\author{M. A. Lara-L\'opez\inst{1}\fnmsep\thanks{\email{mlopez@aao.gov.au}\newline}, 
%Example 
%for footnote, note the usage of the \texttt{fnmsep}
%command as separator between institute number and footnote mark} 
 A. M. Hopkins\inst{1}, A. Robotham\inst{2}, M. S. Owers\inst{1}, M. Colless\inst{1}, S. Brough\inst{1}, P. Norberg\inst{3}, O. Steele\inst{4}, E. N. Taylor\inst{5,6}, D. Thomas\inst{4}
}
\titlerunning{Instructions for authors}
\authorrunning{Lara-L\'opez et al.}
\institute{
Australian Astronomical Observatory, PO Box 915, North Ryde, NSW 1670, Australia
\and 
International Centre for Radio Astronomy Research, The University of Western Australia, 35 Stirling Highway, Crawley, WA 6009,Australia
\and 
Institute for Computational Cosmology, Department of Physics,Durham University, South Road, Durham DH1 3LE, UK.
\and
Institute of Cosmology and Gravitation, University of Portsmouth, Dennis Sciama Building, Burnaby Road, Portsmouth PO1 3FX, UK
\and
Sydney Institute for Astronomy (SIfA), School of Physics, University of Sydney, NSW 2006, Australia
\and
School of Physics, The University of Melbourne, Parkville, VIC 3010, Australia}

\received{...}
\accepted{...}
\publonline{later}


\abstract{%
The stellar mass and metallicity are among the fundamental parameters of galaxies. An understanding of the interplay between those properties as well as their environmental dependence will give us a general picture of the physics and feedback processes ongoing in groups of galaxies. We study the relationships and environmental dependencies between the stellar mass, and gas metallicity for more than 1900 galaxies in groups up to redshift 0.35 using the Galaxy And Mass Assembly (GAMA) survey. Using a control sample of more than 28,000 star-forming field galaxies, we find evidence for a decrement of the gas metallicity for galaxies in groups.
}
\maketitle

\section{Introduction}

The interplay between the stellar mass (\Ms) and gas metallicity  ($Z$) in star forming (SF) galaxies is shown to have a very strong correlation, with massive galaxies showing higher metallicities than less massive galaxies, as quantfied in the M-Z relation (e.g. Lequeux et al. 1979, Tremonti et al. 2004). The M-Z relation has been extensively studied at local redshifts (e.g. Tremonti et al. 04, Kewley \& Ellison 2008), and metallicity has been shown to evolve to lower values even out to relatively low redshifts of $z \approx$ 0.4 (Lara-L\'opez et al. 2009a,b, 2010a, Pilyugin \& Thuan 2011). As redshift increases, the metallicity evolution is stronger.  All studies of the  M-Z relation in high-redshift ($z\sim$ 0.7) galaxies have shown a decrease in metallicity with respect to local galaxies (e.g. Savaglio et al. 2005, Maier et al. 2005, Hammer et al. 2005, Liang et al. 2006, Rodrigues et al. 2008). At redshift $z\sim$2.2, Erb et al. (2006) found that galaxies have a lower metallicity by $\sim$ 0.3 dex, while at redshift $z\sim$3.5, Maiolino et al. (2008)  reported a strong metallicity evolution, suggesting that this redshift corresponds to an epoch of major star-formation activity.

Since metallicity is sensitive to metal losses due to stellar winds, supernovae, and active galactic nuclei (AGN) feedback, the M-Z relation provides essential insight into galaxy formation and evolution. The environment also plays an important role in the gas metallicity of galaxies. Galaxy interactions and mergers can cause gas inflows, morphological transformations, trigger star formation and even lead to activity in the galactic nucleus (Barton, Geller \& Kenyon 2000, Lambas et al. 2003, Nikolic, Cullen \& Alexander 2004, Alonso et al. 2007, Woods \& Geller 2007, Ellison et al. 2008). 

Studies of  galaxies in  pairs and clusters have revealed the environmental effects on the M-Z relation. Kewley, Geller \& Barton (2006) and Ellison et al. (2008) found that galaxies in close pairs are more metal poor by approximately $\sim$0.1 dex at a given luminosity compared with galaxies with no near companion. On the other hand, Ellison et al. (2009) found that galaxies in clusters tend to have higher metallicities by up to $\sim$0.04 dex when compared to a control sample of the same mass, redshift, fibre covering fraction and rest-frame $g-r$ color. This last study emphasizes that the metal enhancements are driven by local overdensities and not simply cluster membership.

This paper introduces  the M-Z relation for galaxies in groups in the GAMA survey. An upcoming paper will present a detailed study of the relationships of the mass, metallicity, SFR, and specific SFR, as well as the Fundamental Plane (Lara-L\'opez et al. 2010b, 2012) for GAMA galaxies in groups. In \S\,  \ref{SampleSelection} we detail the data used for this study, and in \S\,  \ref{MZgama} we introduce the M-Z relation for GAMA. Finally, in \S\,  \ref{Conclusion} we present a summary of our findings.

\section[]{Sample selection}\label{SampleSelection}

GAMA is a spectroscopic survey with data taken with the 3.9m Anglo-Australian Telescope (AAT) using the 2dF fibre feed and AAOmega multi-object spectrograph (Sharp et al. 2006), the spectra were taken with  2 arcsec diameter fibres, a spectral coverage from 3700 to 8900 {\AA}, and spectral resolution of 3.2 {\AA}. For futher details see  Baldry et al. (2010),  Robotham et al. (2010), and Driver et al. (2011).
Galaxies in groups and clusters were selected according to Robotham et al (2012) using a Friends-of-Friends (FoF) algorithm, which has been extensively tested on semi-analytic derived mock catalogues (see also Merson et al. 2011), and has been designed to be extremely robust to the effects of outliers and linking errors.

We have used the Gas AND Absorption Line Fitting algorithm (GANDALF, Sarzi et al 2006) to measure emission lines for the GAMA galaxies. GANDALF is a simultaneous emission and absorption line fitting algorithm designed to separate the relative contribution of the stellar continuum and of nebular emission in the spectra of galaxies, while measuring the gas emission and kinematics. Metallicities were estimated  using the GANDALF emission line GAMA catalogue  using the empirical calibration  provided by  Pettini \& Pagel (2004)  between the oxygen abundance and the O3N2 index. Metallicities were recalibrated to the Bayesian system of Tremonti et al. (2004) using the calibrations of Lara-L\'opez et al. (2012, in preparation). Stellar masses were estimated  by  Taylor et al. (2011),  who estimate the stellar mass--to--light ratio ($M_{\ast}$/L) from optical photometry  using  stellar population synthesis models.

We selected star forming galaxies using the BPT diagram (Baldwin, Phillips \& Terlevich et al. 1981) and the criteria of Kauffmann et al. (2003). For reliable metallicity and SFR estimates, we selected galaxies with a signal-to-noise ratio (SNR) of 3 in  {H$\alpha$}, {H$\beta$}, and  [{N\,\textsc{ii}}]. Also, we constructed 16 volume-limited samples by selecting narrow redshift bins of equal absolute Petrosian r-band magnitudes, as described in Foster et al. (2012) and Lara-L\'opez et al. (2012, in preparation). Our final SF sample is of 1900 group galaxies for GAMA.

To detect any possible change in the metallicity of group galaxies, a control sample  was constructed for each volume-limited sample of GAMA field galaxies at the same redshift, r-magnitude, and stellar mass. Galaxies in groups were excluded from the control sample.

\section[]{The M-Z relation for galaxy groups in GAMA}\label{MZgama}

\begin{figure*}
\includegraphics[scale=0.6]{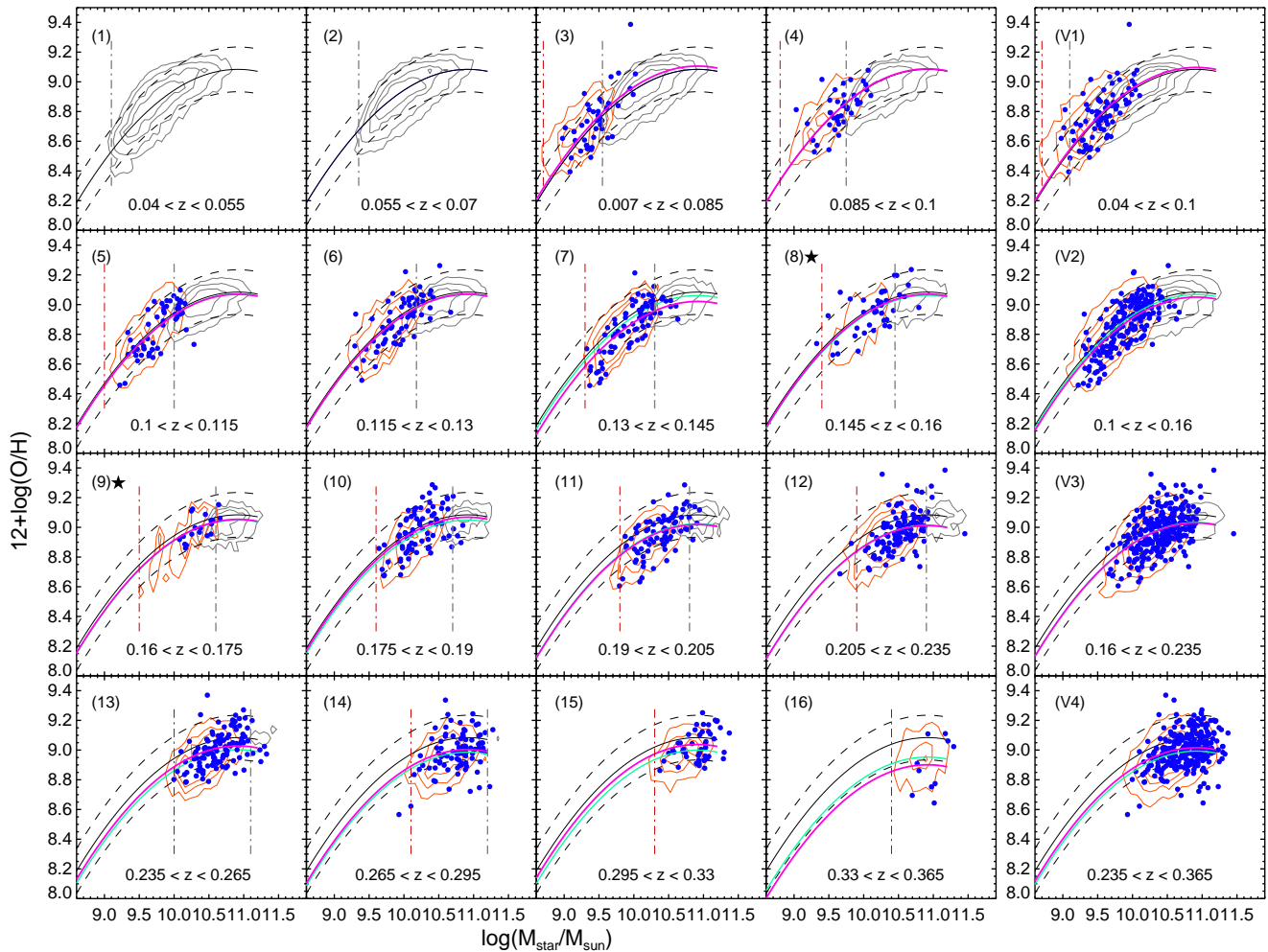}
\caption{M-Z relation for galaxy groups for our 16 volume-limited samples. The right four vertical panels show the concatenated samples V1 to V4 of the redshifts indicated at the bottom of each panel. The star in samples 8 and 9 indicates the GAMA samples for which H$\alpha$ and [NII]~$\lambda$6584  are strongly affected by sky lines. Blue points correspond to our GAMA galaxy group sample. Grey and red contours show field galaxies of SDSS and GAMA, respectively. The black line corresponds to the local M-Z relation of sample V1. The magenta line shows the M-Z relation for each volume-limited sample. The green line shows a fit to the zero point for the galaxy groups taking as a reference the magenta line fit of each panel.}
\label{MZcumulos}
\end{figure*}

The M-Z relation for the 16 volume-limited samples described above is shown in Fig. \ref{MZcumulos}. The blue dots in  Fig. \ref{MZcumulos} correspond to GAMA galaxies in groups. Red contours correspond to our control sample of GAMA  field  SF galaxies, while grey contours represent SDSS galaxies in the same volume-limited sample as described in Lara-L\'opez et al. (2012, in preparation).

According to Lara-L\'opez et al. (2009a,b) and Pilyugin \& Thuan (2011) there are already signs of metallicity evolution for galaxies at redshifts lower than $\sim$0.4. To obtain a reliable control sample for the metallicity evolution, it is important to compare the M-Z relation of groups galaxies with the respective M-Z relation of the control sample in the same redshift bin. In this way, the intrinsic metallicity evolution of each redshift bin will be taken into account.

With the above argument in mind, we proceed to fit the M-Z relation of the control sample for each volume-limited sample, obtaining the following expresion $12+{\rm log(O/H)}=a+b\ x-c\ x^2$,  where $a=-10.5796$, $b=3.5987$, $c=0.1646$, and   $x=\log(M_{\star}/M_{\odot})$. Fixing the $b$ and $c$  coefficients of the M-Z relation, we then fit the zero point $a$ for the galaxy groups in each volume-limited sample, shown in green in Fig. \ref{MZcumulos}.

To improve the statistics of our group sample we grouped volume-limited samples that showed similar evolution. In this way we constructed four concatenated samples (V1 to V4) shown in the right side of Fig. \ref{MZcumulos}. We proceed to fit the M-Z relation for the control and group sample as described above. From the concatenated samples, we can see a small decrement in the metallicity of the group sample in the highest redshift bin.

To further appreciate this shift in metallicity for group galaxies, we constructed metallicity histograms for the control and group samples V1 to V4, shown in Fig. 2. The metallicity of low redshift galaxies for our control and group sample remain mostly the same. For higher redshift samples however, we see that the metallicty of group galaxies decreases by $\sim$0.04 dex. The median metallicity error for group galaxies of samples V1 to V4 is ~0.03 for each one. Although the metallicity error is of the order of the decrement found, it is noteworthy that with a statistically robust sample it is unlikely that the whole population would suffer a similar decrement in the same direction. Therefore, for a statistically robust sample, even small differences could be significant.

% where the highest difference in metallicity is of $\sim$0.04 dex and is shown in our higher redshift sample.

It is important to take into account that each of the samples V1 to V4 are mapping different ranges of stellar mass. Therefore, for low redshift galaxies at low stellar masses, our data suggests that group galaxies do not show any difference in metallicity with respect to field galaxies. On the other hand, massive galaxies in groups at higher redshifts show a decrement in their amount of metals with respect to field galaxies.

\section{Summary and Conclusions}\label{Conclusion}

Using a statistically significant sample of group galaxies we analyzed the M-Z relation up to z$\sim$0.36. Group galaxies at low redshift do not show any significant difference in metallicity when compared with our control sample. On the other hand, we find a decrement in metallicity for 3 of our 4 redshifts bins (V2, V3, V4), where the strongest decrement in metallicity of $\sim$0.04 is shown in the highest redshifts bin V4 (0.235 $<$ z $<$ 0.365). Galaxies in this redshift sample correspond to massive galaxies.

There is a direct correspondence between the stellar mass of individual galaxies and the density of their host group, which means that it is more likely to find low mass galaxies in low density group environments, and vice versa. Therefore, denser environments have a stronger effect on the  metallicity of galaxies, suggesting that in these environments more active gas inflow/outflow processes occur between galaxies and the intra group medium.

Our results agree with Cooper et al. (2008) and Mouchine, Baldry \& Bamford (2007), who used  SDSS galaxies to find that the metallicity of galaxies in the richest environments are only $\sim$0.05 dex higher than those in the poorest environments.The fact that we find a lower metallicity for galaxies in denser environments suggests that at earlier epochs there was a pristine gas interchange that decreased the metallicity of galaxies.

On the other hand, using SDSS data Ellison et al. (2009) found that galaxies in clusters tend to have higher metallicities, however, the SDSS survey is complete for redshifts lower than 0.1, which makes the result of Ellison et al. (2009) suitable for local galaxies. This difference in redshift and selection criteria between our study and Ellison et al. (2009) might be the source of difference between both results. A detailed discussion will be given in Lara-L\'opez et al. (2012, in preparation).

% On the contrary, previous studies have found an increase in metallicity for cluster galaxies at local redshifts (e.g. Ellison et al. 2008). Nevertheless, different results not necessary disagree. In the context

\begin{figure}
\begin{center}
\includegraphics[scale=0.6]{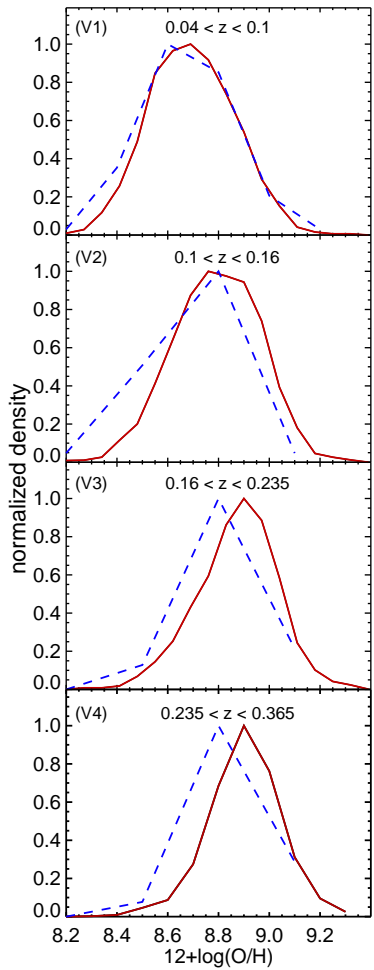}
\caption{Histogram showing the metallicity distribution for samples V1 to V4 of Fig.\ref{MZcumulos}. The red and blue lines correspond to our control and group sample, respectively.}
\end{center}
\label{HistZ}
\end{figure}

\acknowledgements

GAMA is a joint European-Australasian project based around a spectroscopic campaign using the Anglo-Australian Telescope. The GAMA input catalogue is based on data taken from the Sloan Digital Sky Survey and the UKIRT Infrared Deep Sky Survey. Complementary imaging of the GAMA regions is being obtained by a number of independent survey programs including GALEX MIS, VST KIDS, VISTA VIKING, WISE, Herschel-ATLAS, GMRT and ASKAP providing UV to radio coverage. GAMA is funded by the STFC (UK), the ARC (Australia), the AAO, and the participating institutions. The GAMA website is http://www.gama-survey.org/. M. A. Lara-L\'opez thanks to the ARC for a super science fellowship.

% \newpage%%%%%%%%%%%%%%%%%%%%%%%%%%%%%%%%%%%%%%%%%%%%%%%%%%%%%%

\end{document}